\documentclass[aps,prl,preprint,groupedaddress]{revtex4-2}
\usepackage{graphicx}
\graphicspath{{figures/}}
\usepackage{subcaption}
\usepackage{amsmath}
\usepackage{amssymb}
\def\mathbi#1{\textbf{\em #1}}

\begin{document}


\title{Finding Closure Terms Directly from Coarse Data for {2D} Turbulent Flow}


\author{Xianyang Chen}
\author{Jiacai Lu}
\author{Gr\'etar Tryggvason}
\affiliation{Department of Mechanical Engineering \\ Johns Hopkins University, MD,  USA}


\date{\today}

\begin{abstract}
Machine learning is used to develop closure terms for coarse grained model of two-dimensional turbulent flow directly from the coarse grained data by adding a source term to the Navier-Stokes equations to ensure that the coarse-grained flow evolves in the correct way. The source term is related to the average flow using a Neural Network with a relatively simple structure and smoothed slightly to prevent instabilities in a posteriori test. The time dependent coarse grained flow field is generated by filtering fully resolved results and the predicted coarse field evolution agrees well with the filtered results, both for the flow used to learn the closure terms and for flows not used for the learning.
\end{abstract}


\maketitle


\section{Introduction}

Direct Numerical Simulations (DNS) where every spatial and temporal scales in turbulent flows are fully resolved usually require fine grids and long computational times and are therefore too computationally intensive for most flows of industrial interest. Thus, there is a need for reduced order representations of the flow such as Large Eddy Simulation (LES), which are able to predict some of the statistics by evolving only the large scales, using subgrid models to account for the unresolved motion. Generating appropriate subgrid models is still an active area of research and often models perform well for one specific flow but not for others. The classical Smagorinsky model, for example, is too dissipative in wall-bounded flows and prevents laminar flows from transitioning to turbulence (\cite{piomelli1991large}). DNS  that fully resolve all spatial and temporal scales for moderate Reynolds number flows provides enormous amount of data which has been widely used for the development of subgrid models, usually by adjusting coefficients in relationships between the subgrid stresses and the coarse flow, proposed by scaling and other considerations. The recent availability of methods to ``automatically’’ extract correlations from very large data-sets are offering new ways to relate subgrid stresses to the resolved coarse flow. While current focus is mostly on single phase flow, this should be particularly useful for more complex flows, where the details of the subgrid scales are less well known.

In LES the Navier-Stokes equations are filtered, which results in an additional term $\mathcal{F}=\nabla \cdot {\boldsymbol \tau}_{SGS}$, where the subgrid stresses are related to the filtered fully resolved flow by ${\boldsymbol \tau}_{SGS} = \widetilde {{\bf u} {\bf u} } - \tilde {\bf u} \tilde {\bf u}$. Here, $ \tilde {\bf u} $ is the filtered velocity and $\widetilde {{\bf u} {\bf u} }$ is the filtered product. Comparisons of ${\boldsymbol \tau}_{SGS}$ computed from a model and computed from a filtered fully resolved field are generally referred to as a priori tests. Here, instead of using the filtered fully resolved solution, we compute $\mathcal{F}$ so that the time dependent coarse flow evolves correctly. The coarse flow is found by filtering a fully resolved one, but otherwise it is not used (except for comparisons) and in principle the coarse flow could come from other sources, such as experimental data, or from a field coarsened in such a way that the connections with the filtered governing equations is not obvious.

A number of researchers have recently explored the use of machine learning to develop reduced order models in fluid mechanics (\cite{ling2016reynolds},\cite{xiao2016quantifying}, \cite{wang2017physics}, \cite{wu2019reynolds}, \cite{tracey2015machine}, \cite{weatheritt2017hybrid}, \cite{ma2015using}, \cite{ma2016using}). See \cite{duraisamy2019turbulence} for a review. Most of the contributions have focused on closure models for Reynolds-Averaged Navier-Stokes (RANS) equations, but some efforts have also been made for LES. \cite{sarghini2003neural} used a Neural Network (NN) to predict the turbulent viscosity coefficient for the mixed model in a turbulent channel flow. \cite{gamahara2017searching} modeled the subgrid stress tensor directly using NN for turbulent channel flows, without making any assumption about the model form, but the results did not show obvious advantages over the Smagorinsky model. \cite{beck2019deep} reported instabilities in a posteriori test with closure models using Residual Neural Network (RNN), even though a priori test showed a high correlation coefficient. Predicting the turbulent viscosity instead of the subgrid stresses did, however, eliminate the instabilities.  \cite{wang2018investigations} included second derivatives of the velocity field as inputs for NN to predicted subgrid stresses, showing a better result than the classical Smagorinsky and the dynamic Smagorinsky model for the energy dissipation. \cite{xie2020modeling} used NN to model the source term, rather than the subgrid stresses in homogeneous isotropic turbulence and while they found a high correlation coefficients (0.99), they had to add artificial diffusion to damp out instabilities. They also extended their modeling to turbulence in compressible flows in \cite{xie2020spatially}. We note that while \cite{xie2020modeling} work with the source term, they compute it from the subgrid stresses, rather than finding it directly from the coarse field, as we do. For 2D turbulence, \cite{maulik2017neural} presented a priori test using NN models for deconvolution of the flow field, as well as for 3D homogeneous isotropic turbulence, and turbulence in stratified flow. In \cite{maulik2019subgrid}, they extended their work by conducting a posteriori test for 2D turbulence, but had to truncate the predicted subgrid source terms to supress instabilities.

\section{Method}

\subsection{The coarse field}

We conduct fully resolved simulations of unsteady two dimensional flow and filter the results to produce a coarse grained velocity and pressure fields. The simulations are done using a standard projection method on a regular staggered grid. A second order Runge-Kutta method is used for the time integration, a second order Quick scheme is used for the advection terms and second order centered approximations are used for all other spatial derivatives. The coarse grained field, denoted by $\mathbf{\tilde{u}}$, is generated by spatial filtering. Here we use a Gaussian: 
\begin{equation}
    G_{\Delta}=(\frac{6}{\pi\Delta^2})^{3/2}exp(\frac{-6\|\mathbf{x}-\mathbf{x'}\|^2}{\Delta^2}),
    \label{filter}
\end{equation}
where $\Delta$ is the cutoff length that separates the length scales.

To evolve the coarse field, we add a source term $\mathcal{F}$ to the Navier-Stokes equations:
\begin{equation}
    \frac{\partial \mathbf{\tilde{u}}}{\partial t}+ \nabla \cdot \mathbf{\tilde{u}}\mathbf{\tilde{u}}=-\frac{1}{\rho}\nabla \tilde{p}+\nu \nabla^2 \mathbf{\tilde{u}}+\mathcal{F}.
\label{NSeq}
\end{equation}
We then estimate the time derivative  by finding the difference in the velocity at two-time levels and compute the source terms by:
\begin{equation}
\mathcal{F}=\frac{\partial \mathbf{\tilde{u}}}{\partial t}+ \nabla \cdot \mathbf{\tilde{u}}\mathbf{\tilde{u}}+\frac{1}{\rho}\nabla \tilde{p}-\nu \nabla^2 \mathbf{\tilde{u}},
\label{source}
\end{equation}
where $\tilde{p}$ is the filtered pressure. We note that this is different from what is usually done for LES modeling using DNS results, where the subgrid stresses are computed from the fully resolved results. While we have the fully resolved flow field and could thus compute the subgrid stresses directly in the usual way, our motivation is to test an approach applicable to cases where we only have the coarse field. The extra term,  $\mathcal{F}$, is intended to ensure that equation (\ref{NSeq}) evolves the coarse field correctly.

\subsection{Neural Network (NN) Architecture}

The key assumption in LES is that the subgrid stresses can be related to the coarse or large scale flow and we assume that the same is true for the source term (even if it is not computed directly as the divergence of the stresses). The source term cannot depend directly on the velocities since we can always add a constant translation, so it must depend on the derivatives of the velocity. For three-dimensional flows, \cite{wang2018investigations} examined the relevance of a number of variables as input for neural network for the subgrid stresses and determined that the first and second derivative of the velocity were most important. \cite{xie2020modeling} determined the source term $\nabla \cdot \tau_{SGS}$ (which they refer to as the SGS force) using the first derivatives, but included several surrounding spatial points, instead of just the single point under consideration. They found that working with the full source term, rather than the stresses, resulted in a much better agreement with the filtered data. Here, we assume that the source term depends on the local value of the first and second derivatives: $ \frac{\partial u}{\partial x} $, $ \frac{\partial u}{\partial y} $, $ \frac{\partial v}{\partial x} $, $ \frac{\partial^2 u}{\partial x^2} $, $ \frac{\partial^2 u}{\partial x \partial y} $, $ \frac{\partial^2 u}{\partial y^2} $ and $ \frac{\partial^2 v}{\partial x^2} $. Only 7 components are needed because of the incompressibility condition. 
Because the flow is homogeneous and without any directional preference, we can use the same expression for the two components $\mathcal{F}_x$  and $\mathcal{F}_y$,
since by rotating the coordinate system by $90^{\circ}$ in counterclockwise direction, the new ``x'' axis collapses with original ``y'' axis. Thus we can use the same mapping function for the $y$ component, by slightly ``rotating'' the inputs. Thus
\begin{eqnarray}
\mathcal{F}_x =\mathcal{M} \Biggl( \frac{\partial u}{\partial x}, \frac{\partial u}{\partial y},  \frac{\partial v}{\partial x}, \frac{\partial^2 u}{\partial x^2}, \frac{\partial^2 u}{\partial x \partial y}, \frac{\partial^2 u}{\partial y^2},\frac{\partial^2 v}{\partial x^2}  \Biggr), \\
\mathcal{F}_y =\mathcal{M} \Biggl(\frac{\partial v}{\partial y}, -\frac{\partial v}{\partial x},  -\frac{\partial u}{\partial y}, \frac{\partial^2 v}{\partial y^2}, -\frac{\partial^2 v}{\partial x \partial y}, \frac{\partial^2 v}{\partial x^2},-\frac{\partial^2 u}{\partial y^2}  \Biggr),
 \label{nnmapping}
\end{eqnarray}
where $\mathcal{M}$ denotes the function found by the training. 
In this way we preserve the rotational invariance property, are able to use the data for the two components together in one big dataset, and train only once.

\begin{figure}[t]
	\centering
    \includegraphics[height=3in]{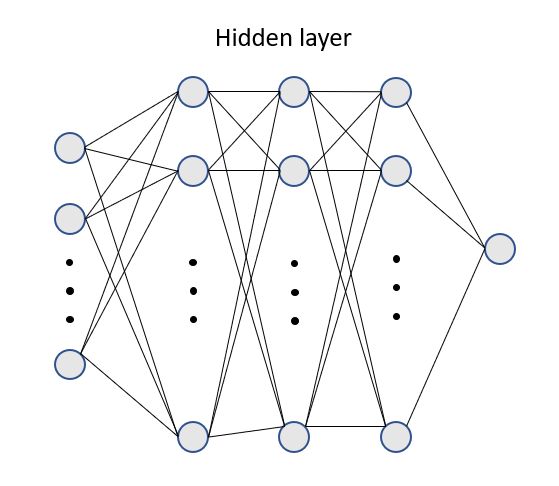}
    \caption{The structure of the Neural Network.}
    \label{NN}	
\end{figure}

\begin{figure}
    \begin{subfigure}[h]{0.5\linewidth}
    	\centering
    	\includegraphics[height=2in]{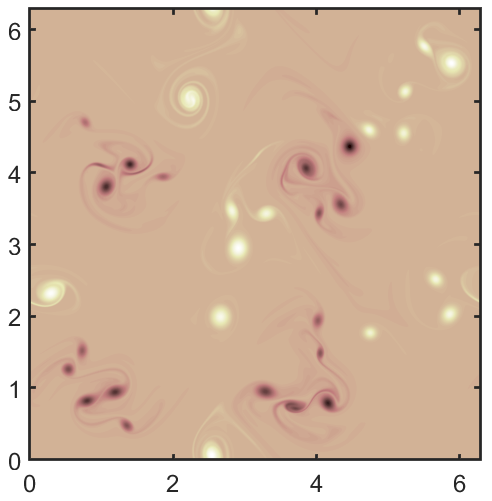}
    	\caption{$t^*=4.55$}
    \end{subfigure}%
   \begin{subfigure}{0.5\linewidth}
    	\centering
    	\includegraphics[height=2in]{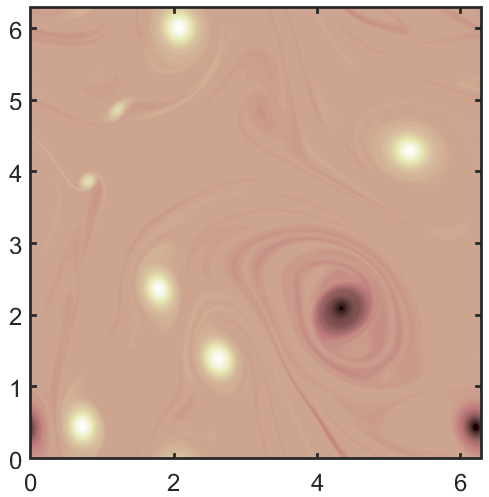}
    	\caption{$t^*=22.73$}
    \end{subfigure}%
    \caption{DNS vorticity at nondimensional times $t^*=4.55$ and $22.73 $  (scaled by $\tau_L$).}
    \label{DNSvt}
\end{figure}

The Neural Network is a multilayered structure, with a linear vector transformation followed by a nonlinear activation function at each layer, achieving complicated regression from inputs to outputs. Here, we use a Neural Network with a relatively simple structure, containing one input layer with 7 neurons, 3 hidden layers with 20 neurons in each layer, and an output layer with 1 neuron. Neurons in adjacent layers are fully connected to each other. Figure \ref{NN} illustrates the structure of the Network. The transformation from layer $l$ to layer $l+1$ can be expressed as
\begin{equation}
    \mathbi{Y}^{l+1}=\sigma(\mathbi{W}^l\mathbi{X}^{l}+\mathbi{b}^l),
\end{equation}
where $\mathbi{X}\in\mathbb{R}^{m\times 1}$ is the data in layer $l$, $\mathbi{Y}\in\mathbb{R}^{n\times 1}$ is the data in layer $l+1$, $\mathbi{W}\in\mathbb{R}^{n\times m}$ and $\mathbi{b}\in\mathbb{R}^{n\times 1}$ represents the weighting matrix and the bias term that needs to be optimized. $\sigma(x)$ stands for the RELU activation function, $\sigma(x)=max(x,0)$, used to introduce nonlinearity into the mapping. For effective learning, the input data has been preprocessed in such a way that $\mathbi{X}^0=\frac{1}{\alpha}(\mathbi{X}-\mathbi{X}_{mean})$, where $\mathbi{X}_{mean}$ and $\alpha$ are the mean and standard deviation of the raw data. The NN structure is built with Keras in Python 3.7 and the mapping is optimized by stochastic gradient descent method by monitoring the mean square error as the loss function. Based on our experiment, 500 epochs are generally enough for finding the optimum and the loss function does not decrease significantly with a further increase in the number of epochs.

\section{Results and Discussion}

\begin{figure}
	\centering
	\begin{minipage}{.48\textwidth}
		\centering
		\includegraphics[height=2.3in]{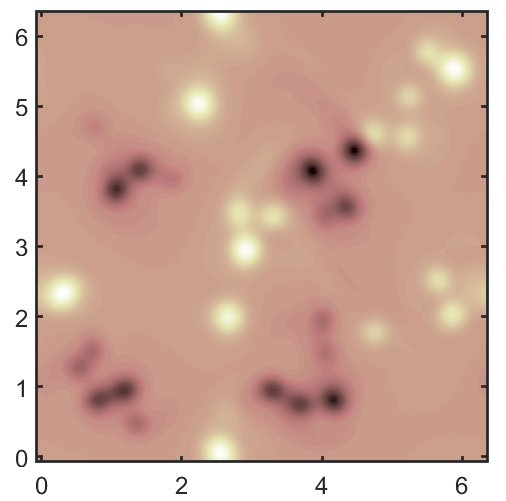}
		\captionof{figure}{Filtered vorticity field with filter width $\Delta=0.4$ at $t^*=4.55$.}
		\label{fdns}
	\end{minipage}\hfill
	\begin{minipage}{.48\textwidth}
		\centering
		\includegraphics[height=2.5in]{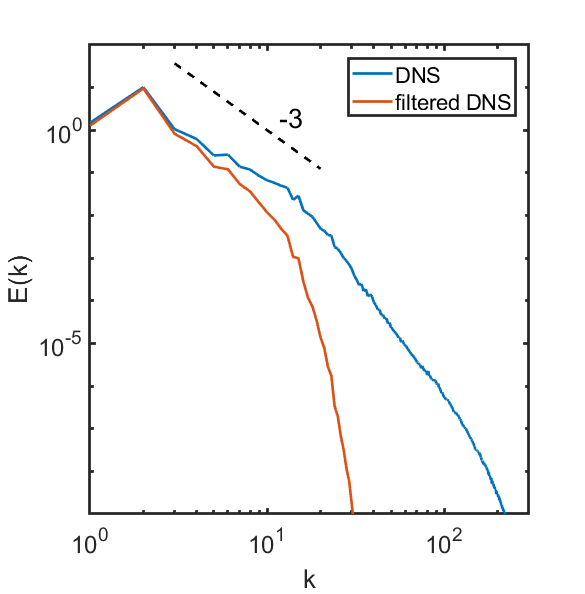}
		\captionof{figure}{Energy spectrum for the DNS field and the filtered DNS field at $t^*=4.55$.}
		\label{fspectrum}
	\end{minipage}	
\end{figure}

To generate a 2D turbulent flow in a doubly periodic domain, we start with four perturbed shear layers that quickly break up by Kelvin-Helmholtz instability into approximately isotropic flow as seen in figure 2, where the vorticity is shown at two times. The domain size is $2 \pi \times 2 \pi$, resolved by a $2048 \times 2048$ grid. Due to the inverse energy cascade characteristic of 2D turbulence, small vortices grab each other and form larger ones, as shown in the evolution of the vorticity field. In this case, turbulence energy dissipation rate is $ \epsilon = 2\nu(\overline{S'_{ij}}    \overline{S'_{ij}})=0.179 $  and the Kolmogorov length scale is approximately $ \eta_k =  (\nu^3/\epsilon)^{1/4} =0.009 $. The Reynolds number defined by the integral length scale is $ Re_{L}=u_{RMS}L/\nu=2335 $, where $ u_{RMS}$ is the RMS velocity $ u_{RMS}=\sqrt{E_{k}}=3.28 $ and the integral length scale can be derived from that $ L=({\pi/}{2u_{RMS}^2}) \int\frac{E(k)}{k}\,dk=0.713 $. The large eddy turnover time scale is $ \tau_L=L/u_{RMS}=0.22 $, and the simulation is run for over 45 large eddy turnover times.

\begin{figure}
	\centering
    \includegraphics[height=2.5in]{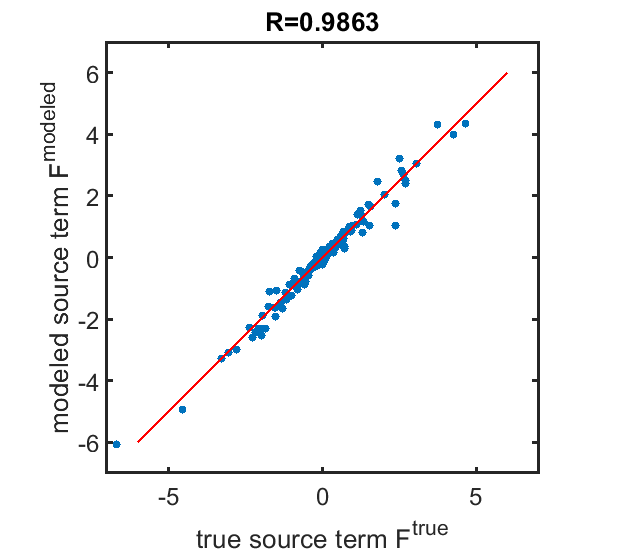}
    \caption{The predicted value of the component of the source term versus the true source terms. The correlation coefficients are shown on the top of the frame. The $45^{\circ}$ red line shows a perfect fit.}
    \label{correlation}
\end{figure}

\subsection{NN learning}

To generate a coarse flow field, we apply a Gaussian filter to the DNS flow field so that the original $2048\times 2048$ field is coarsened to $128 \times 128$ grid points, or $\Delta_c=16\Delta_f$, where $\Delta_c$ and $\Delta_f$ denote the coarse grid spacing and the fine grid spacing, respectively. The filter width in equation (\ref{filter}) is chosen to be $\Delta=0.4$, which is about half the integral length scale $L $ and much larger than the grid spacing of the coarse mesh ($\Delta_c=0.049$). Note that the filter applied here is more aggressive than that in \cite{maulik2017neural}, who used $\Delta_c=8\Delta_f$. The filter size is also much larger than the fine grid spacing ($\Delta=130\Delta_f$), and more aggressive than in \cite{xie2020modeling} where $\Delta=32\Delta_f$. Figure \ref{fdns} shows the filtered vorticity field on the $128 \times 128$ grid. We obtain the 1D energy spectrum $ E(k)=\frac{1}{2}|\hat{u'}(k)|^2 $, where $ \hat{u'}(k)$ represents the Fourier Transform of the velocity fluctuation field, by taking the Fourier transform of the fluctuation along the x direction and average in the y direction. Doing it in a reversed way (taking the Fourier transform along the y direction and averaging in the x direction) gives similar results. The energy spectrum for the fully resolved DNS solution and the filtered flow are shown in figure \ref{fspectrum}. The spectrum shows the classical -3 power law in the inertial range for 2D turbulence (\cite{kraichnan1967inertial}). 

Here, we sample the data every $\Delta t^*=2.27$ from $t^*=2.27$ to $t^*=27.27$ for the Neural Network training, giving a total of 12 filtered DNS fields, each containing $128\times 128$ data points (total of $393,216$ samples by combining data for both components). Before the data was used to train the NN, it was normalized by subtracting the mean and dividing by the standard deviation, enforcing a mean and standard deviation of 0 and 1, respectively. The data is split into a training set (90\%) and a validation set (10\%) randomly in order to perform a cross-validation. Figure \ref{correlation} shows the predicted value of one component of the source term versus the true source term computed from the data by equation (\ref{source}). The correlation coefficients are shown on the top of the frame. It achieves a value higher than 0.98 after 500 epochs. Increasing the number of epochs, as well as using different initial guesses, has minimal impact on the results. The high correlation coefficient in the cross-validation test is about the same as \cite{xie2020modeling} found and higher than what \cite{wang2018investigations} found (0.7).

\begin{figure}[t]
	\centering
	\begin{subfigure}[ht]{0.29\textwidth}
		\includegraphics[height=1.6in]{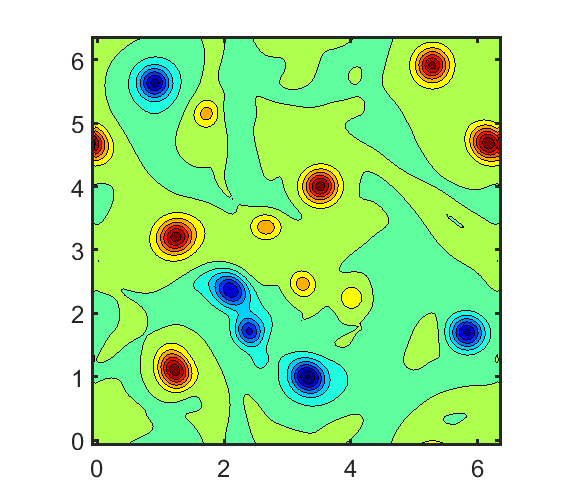}
	\end{subfigure}
	\begin{subfigure}[ht]{0.29\textwidth}
		\includegraphics[height=1.6in]{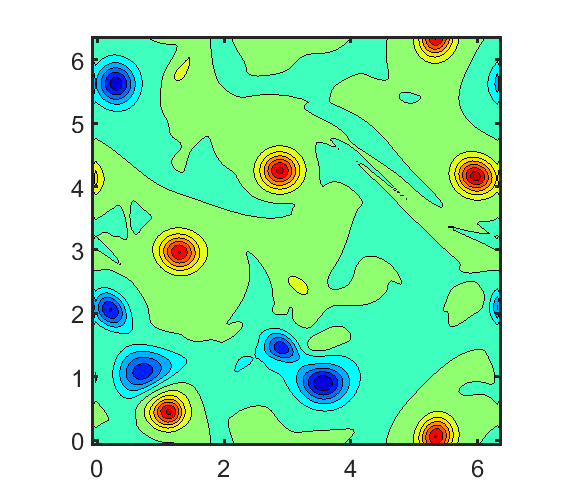}
	\end{subfigure}
	\begin{subfigure}[ht]{0.36\textwidth}
		\includegraphics[height=1.6in]{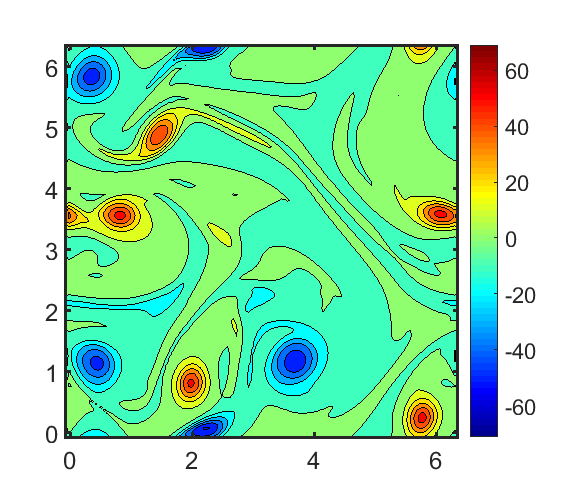}
	\end{subfigure}
	\caption{The vorticity contours at $t^*=11.36$. (a) Filtered DNS vorticity field; (b) LES vorticity field with NN models; (c) Vorticity field on coarse grid with no model.}
	\label{vtcomparison}
\end{figure}

The source term coming from the neural network is not a completely smooth function and the noise is sometimes amplified when used in the modified Navier-Stokes equations. To eliminate that we smooth the source term slightly, using a box filter to smooth it over a 5 by 5 stencil on the course grid. Using a Gaussian filter with the same cut-off length as used to coarsen the fully resolved flow also gives similar results. We note that instabilities were also seen by \cite{maulik2019subgrid}, \cite{xie2020modeling} and \cite{beck2019deep}. In order to avoid the numerical instabilities for 2D flows, \cite{maulik2019subgrid} forced the subgrid stress to be only dissipative at large scales by truncating the source term, assuming that the small scales only cause dissipation of the kinetic energy. \cite{piomelli1991subgrid} pointed out that forward- and backscatter are present in approximately equal amounts in 3D. This was also shown by \cite{maulik2019subgrid}, who found that even in 2D flows about half of the grid points transfer energy from fine scales to large scales. Here, we avoid excluding the backscatter mechanism. \cite{xie2020modeling} prevented instabilities in 3D by including artificial dissipation into the momentum equation to damp out the high frequency fluctuations.

\subsection{A posteriori tests}

Studies of subgrid stresses in LES typically distinguish between a priori comparison of predicted subgrid stresses with subgrid stresses computed by filtering fully resolves results, and a posteriori comparison of how well the flow field computed by LES matches the filtered fully resolved results. A somewhat disappointing finding is that a successful a priori test does not guarantee a good a posteriori comparison. In our case, where the closure terms are computed directly from the coarse grained result, a priori test is irrelevant. We are not trying to reproduce the subgrid stresses or the source term computed from the filtered DNS solution, but to ensure that the coarsened field evolves correctly. The a posteriori test on the other hand, where we compare the evolution of the coarse grained field to the evolution of the filtered fully resolved field, is of course the ultimate test.

\subsubsection{The training data}
The purpose of a subgrid model is to ensure that the coarse field reproduces the evolution of the fully resolved field, filtered at each instance in time. Since the flow is highly unsteady and slightly different initial conditions diverge at long times, we do not expect the coarse solution to follow the filtered field exactly, except initially. Thus, we can only do a detailed comparison between the spatial fields at early times and must be content with comparing statistical quantities at later times. 

We use the filtered DNS field at $t^*=4.55$ (figure \ref{fdns}) as the initial condition for the a posteriori test. The LES uses a $128\times 128$ grid, with the NN predicted source terms $\mathcal{F}^{model}$ added on the right hand side of the momentum equation, followed by a slight spatial smoothing of the modeled source term field. We also ran an under-resolved DNS simulation on a $128\times 128$ grid for comparison, also with the filtered fully resolved solution at $t^*=4.55$ as initial conditions. Figure \ref{vtcomparison} shows the instantaneous vorticity field as found by the filtered DNS, the LES and the under-resolved DNS at $t^*=11.36$. The agreement between the filtered DNS and the LES results is reasonably good and the coarse grid solution (under-resolved DNS) is further away from the filtered solution. This suggests that the NN model captures the evolution of the coarse filtered field, at least for the first $6.81$ large eddy turnover times.  

\begin{figure}[t]
	\centering
	\begin{minipage}[ht]{.48\textwidth}
		\centering
		\includegraphics[height=2.4in]{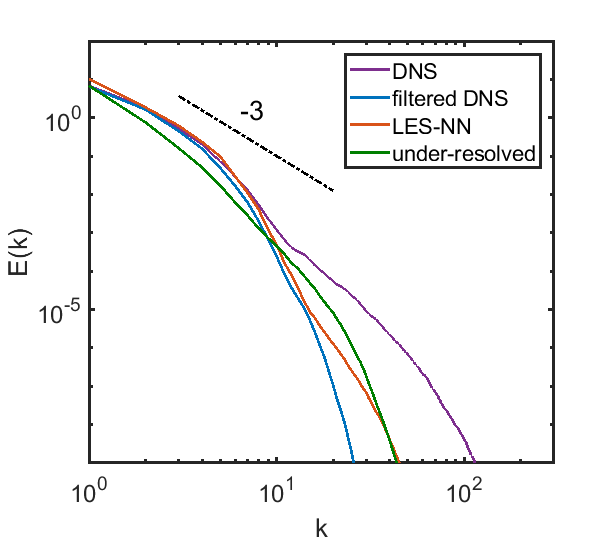}
		\captionof{figure}{The 1D energy spectrum of the DNS field, the filtered DNS field, the LES field and the coarse grid solution.}
		\label{posteriorispectrum}
	\end{minipage}\hfill
	\begin{minipage}[ht]{.48\textwidth}
		\centering
		\includegraphics[height=2.4in]{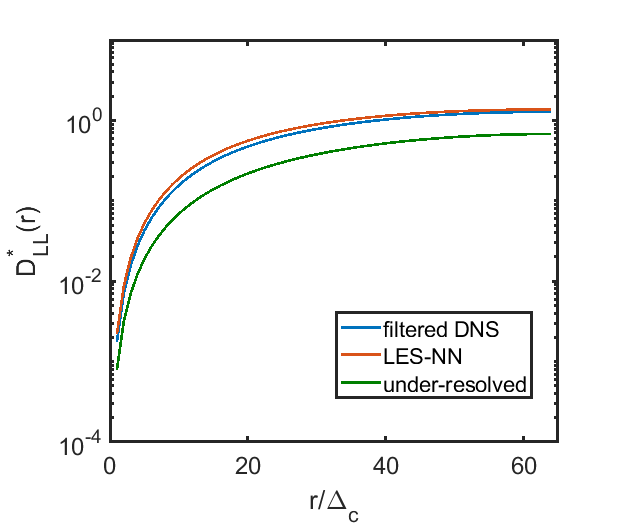}
		\captionof{figure}{The second order nondimentional longitudinal structure function $D^*_{LL}$.}
		\label{posterioristructure}
	\end{minipage}	
\end{figure}

\begin{figure}[t]
	\centering
	\includegraphics[height=2.5in]{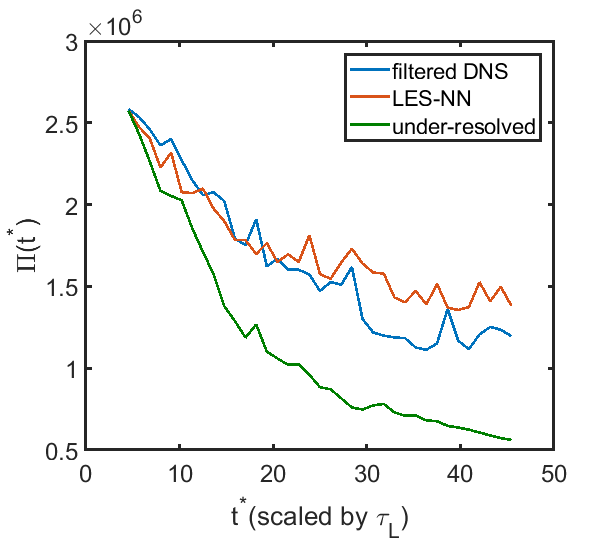}
	\caption{The evolution of the enstropy for the filtered DNS, LES and an under-resolved solution.}
	\label{enstropy1}
\end{figure}

Since it is unrealistic to expect the coarse field obtained by filtering the fully resolved results and the coarse field evolved by the model equations to match completely at long times, we assess the performance of the model by examining if statistical quantities are preserved. Here, we calculate the 1D energy spectrum and the $2^{nd}$ order structure function for the filtered DNS field, LES field with NN model, and the coarse grid solution with no models. The 1D longitudinal structure function is defined as $D_{LL}(r)=\overline{(u' (\mathbf{x}+\mathbf{r})-u' (\mathbf{x}))^2}$, where $u'$ denotes the horizontal velocity fluctuation. We average the field at 11 different times and nondimensionalize the structure function by $u^2_{RMS}$. Averaging with more samples or at other times does not significantly change the structure function or the 1D spectrum. As shown in figure \ref{posteriorispectrum}, the energy spectrum obtained by the NN modeled LES follows the $-3$ power law well in the inertial range. The coarse grid solution is however, a little bit away from the $-3$ power law at large scales. Figure \ref{posterioristructure} shows the 1D $2^{nd}$ order nondimensional longitudinal structure function versus $r/\Delta_c$. The NN modeled LES structure function agrees well with the filtered fully resolved one, and is much closer than the coarse grid solutions. We also plot the enstropy, defined as $\Pi=\oint\omega\cdot\omega\,dS $, versus time for the three different cases in figure \ref{enstropy1}. It can be seen that there is an obvious improvement in matching the enstropy evolution by using the NN modeled LES compared to the under-resolved solution. We note that it takes 25 days on a single Intel core to evolve the DNS solution to $t^*=45.46$ but only less than half an hour for the NN modeled LES.

\subsubsection{Non-training data}

In the previous section we show that the Neural Network closure model works well in terms of matching both the instantaneous flow field at early times as well as the turbulence statistics at long times for the data used to train the model. Here we apply the model to a different case generated by 4 vertical shear layers separating flows with discontinuous velocities, still on a $2048\times 2048$ grid for a domain size of $2\pi \times 2\pi$. The Reynolds number, defined by the integral length scale, is about the same as for the previous case, or $Re_L=2685$. The Kolmogorov length and the RMS velocity fluctuations are $\eta_k=0.0082$ and $u_{RMS}=3.29$. The large eddy turnover time is $\tau_L=0.24$. We choose the same filter size $\Delta=0.4$ and coarsen the DNS fields onto $128\times 128$ grid points, since the NN model has only been trained for one filter size. 

\begin{figure}[ht]
	\centering
	\begin{subfigure}[ht]{0.29\textwidth}
		\includegraphics[height=1.6in]{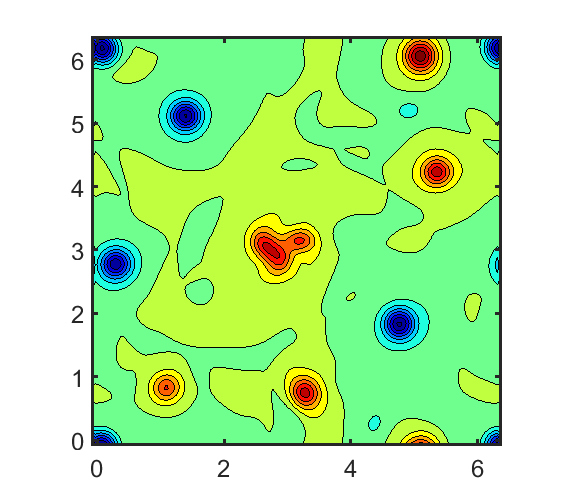}
	\end{subfigure}
	\begin{subfigure}[ht]{0.29\textwidth}
		\includegraphics[height=1.6in]{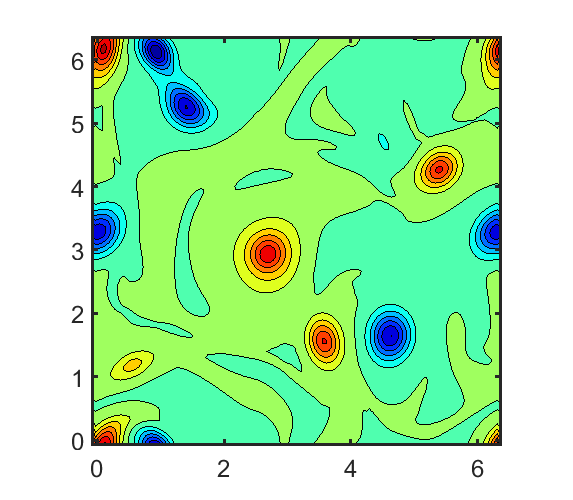}
	\end{subfigure}
	\begin{subfigure}[ht]{0.36\textwidth}
		\includegraphics[height=1.6in]{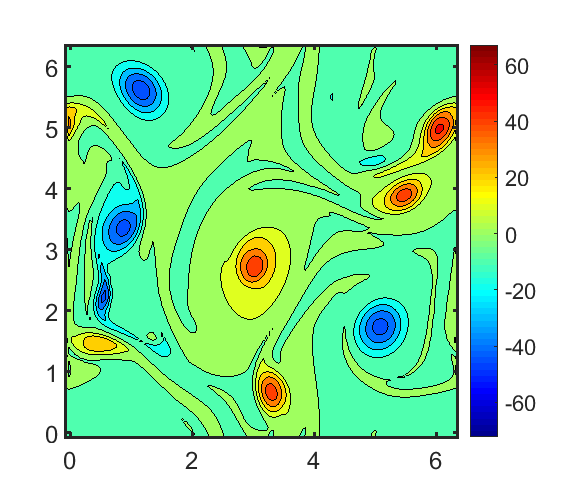}
	\end{subfigure}
	\caption{Contours at $t^*=10.42$ of (a) filtered DNS vorticity field (b) LES vorticity field with NN model (c) under-resolved vorticity field on coarse grid.}
	\label{vtcomparison2}
\end{figure}

\begin{figure}[ht]
	\centering
	\begin{minipage}[ht]{.48\textwidth}
		\centering
		\includegraphics[height=2.4in]{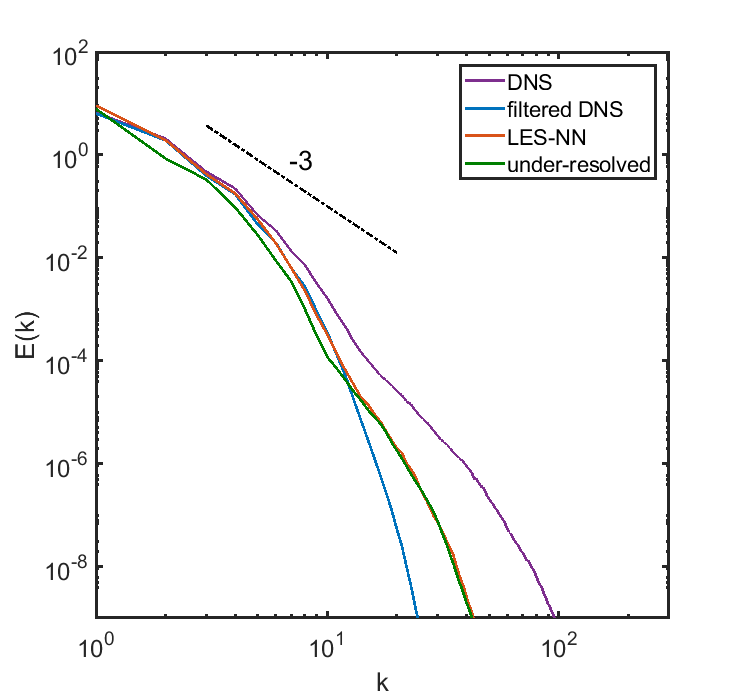}
		\captionof{figure}{The 1D energy spectrum of the DNS field, the filtered DNS field, the LES field and the coarse grid solution.}
		\label{lastspectrum}
	\end{minipage}\hfill
	\begin{minipage}[ht]{.48\textwidth}
		\centering
		\includegraphics[height=2.4in]{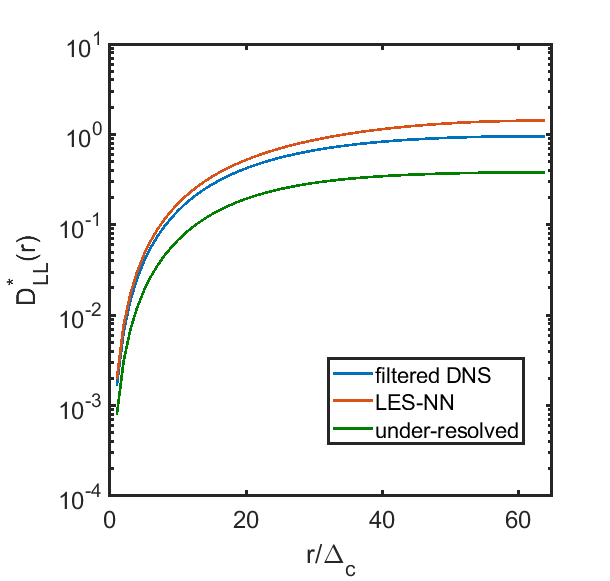}
		\captionof{figure}{The second order nondimentional longitudinal structure function $D^*_{LL}$.}
		\label{laststructure}
	\end{minipage}
	
\end{figure}

\begin{figure}[ht]
	\centering
	\includegraphics[height=2.5in]{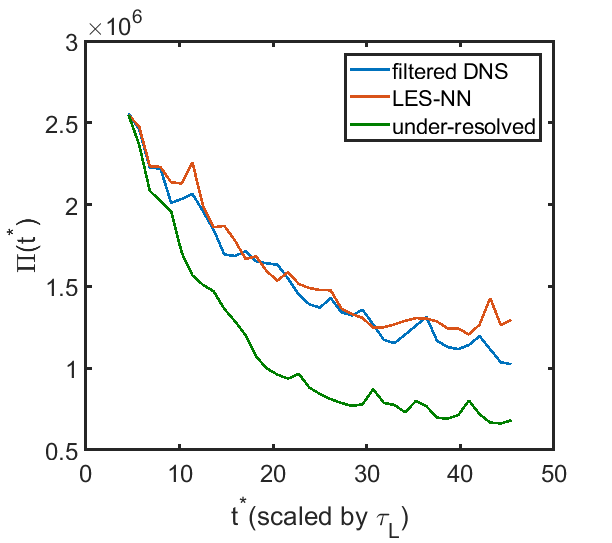}
	\caption{The evolution of the enstropy for the filtered DNS, the LES and the under-resolved solution.}
	\label{enstropy2}
\end{figure}

We note that we keep the majority of the inputs to the NN within the range of the training data, facilitating a nonlinear interpolation of the data. We start the a posteriori test from the filtered DNS field at $t^*=4.17$, and plot the instantaneous vorticity field at $t^*=10.42$ in figure \ref{vtcomparison2}, along with the under-resolved solution on a $128\times 128$ grid. The LES vorticity field shows a better agreement with the true field after nearly 7 large eddy turnover times than the under-resolved one. Figure \ref{lastspectrum} shows that the LES matches the energy spectrum well and is better than under-resolved solution in terms of the structure function, shown in figure \ref{laststructure}, and the evolution of the enstropy in figure \ref{enstropy2}. 


\section{Conclusions}

The main goal of the present paper is to explore how we can generate closure terms for a reduced order, or coarse grained, model directly from the coarse data, by finding what  we need to add to the standard momentum equations so that the coarse flow evolves in the same way as the filtered fully resolved flow. The not unexpected conclusion is that it works, at least for the problem examined here. Here we start with a fully resolved flow field and could, of course, compute the standard subgrid stresses by filtering the fully resolved velocities and the product of the fully resolved velocities separately and subtract the product of the filtered velocities from the filtered product, as is usually done (see \cite{maulik2019subgrid}). However, if the fully resolved solution is not available, or if the coarsening is carried out in such a highly nonlinear way that is not possible to write down the analytical form of the source terms, then the present approach should still work. When the subgrid stresses are available they show up in the averaged momentum equations as fluxes and their divergence ensures that no new momentum is added. This is not preserved when we work with the source term directly, although we expect conservation to be approximately satisfied. While the stresses can, in principle, be recovered by integration at any given time, for flows in fully periodic domains those are only specified up to a time dependent constant, potentially causing difficulties with using stresses from different times for the learning. This problem should not exist in wall-bounded flows, but we have not explored this further at the present time. We also note that we took the subgrid stresses to depend on any coarse quantity we could think of. It is likely that this number can be reduced, but doing so is not needed for the purpose of this study.

\appendix
\counterwithin{figure}{section}
\section*{Appendix: A priori  test}

\setcounter{figure}{0} \renewcommand{\thefigure}{A.\arabic{figure}}

\begin{figure}[ht]
	\centering
	\begin{subfigure}[ht]{1\textwidth}
		\includegraphics[height=2in]{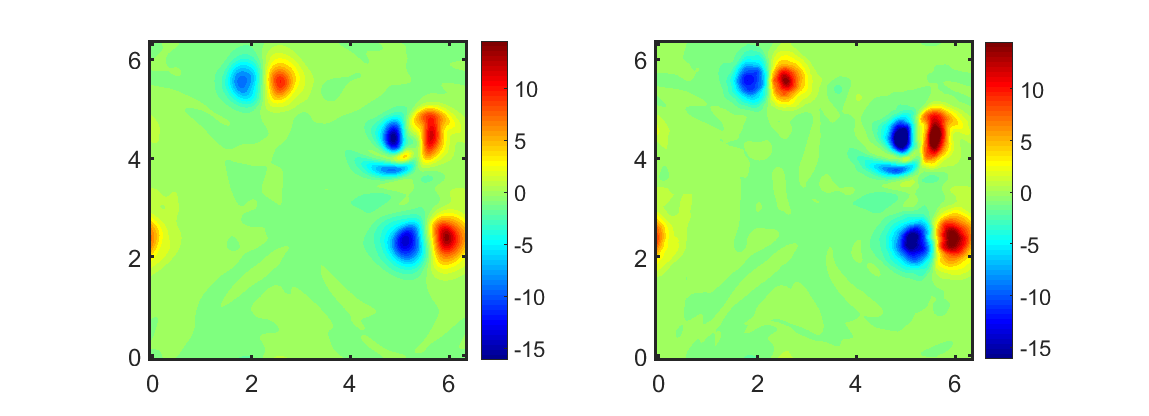}
	\end{subfigure}
	\begin{subfigure}[ht]{1\textwidth}
		\includegraphics[height=2.05in]{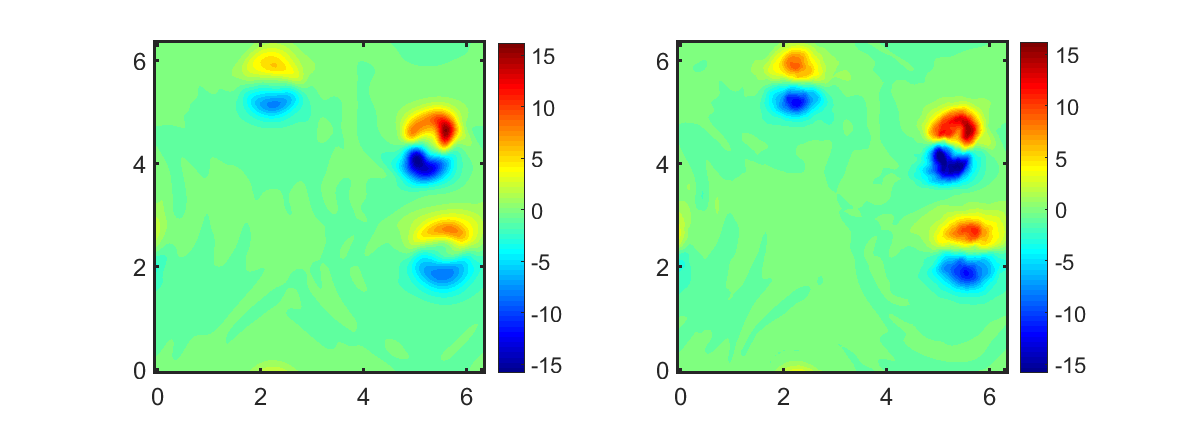}
	\end{subfigure}
    \caption{A comparison of the instantaneous true $\mathcal{F}$ field (left) and the predicted one (right) for both x (top) and y components (bottom) at $t^*=36.36$.}
    \label{divtau}	
\end{figure}

Although a priori tests are generally irrelevant to our approach, where the closure terms are found directly from the coarse  data, 
we expect the source term computed using the coarse field by equation (\ref{source}), to be the same as the one computed from filtering the fully resolved results.
Thus, we show in figure \ref{divtau} a comparison of the source term predicted by the Neural Network and the source term computed from the fully resolved flow at one time for the training data. Note that the second one is not the source term used to train the Neural Network---which is computed from the evolution of the filtered field by equation (\ref{source}). The agreement is reasonably good for both components, with a mean square error (MSE) of 1.103 and 0.467 for the x and the y components respectively. The MSE is defined by $MSE=\frac{1}{N}\sum \| \mathcal{F}^{true}-\mathcal{F}^{modeled}\|^2$. 

\begin{acknowledgments}
This research was supported in part by the National Science Foundation Grant CBET-1953082. We also acknowledge the use of computational resources at the Maryland Advanced Research Computing Center (MARCC).
\end{acknowledgments}


\bibliography{apssamp.bib}
\end{document}